\documentclass[aps,onecolumn,pre]{revtex4}
\usepackage{latexsym, verbatim}
\usepackage{amsmath}
\usepackage{mathrsfs}
\usepackage{amsthm}
\usepackage{amssymb}
\usepackage{dsfont}
\usepackage{fancybox}
\usepackage{bm}
\usepackage{fancyhdr}
\usepackage{subfigure}
\usepackage{graphicx}
\usepackage{color}

\begin{document}

\title{\bf Asymmetric Wave Propagation Through 
Saturable Nonlinear Oligomers}

\author{D. Law}
\affiliation{Department of Mathematics and Statistics,
Amherst College,
Amherst, Massachusetts 01002, USA} 

\author{J. D'Ambroise}
\affiliation{Department of Mathematics and Statistics,
Amherst College,
Amherst, Massachusetts 01002, USA} 

\author{P.G.~Kevrekidis}
\affiliation{Department of Mathematics and Statistics, University of Massachusetts,
Amherst, Massachusetts 01003, USA}

\author{D.\ Kip}
\affiliation{Faculty of Electrical Engineering, Helmut Schmidt University, Hamburg 22043, Germany}

\begin{abstract}
In the present paper we consider nonlinear dimers and trimers 
(more generally, oligomers) embedded within a linear Schr{\"o}dinger
lattice where the nonlinear sites are of saturable type. 
We examine the stationary states of such chains in the form
of plane waves, and analytically compute their reflection and transmission
coefficients through the nonlinear oligomer, as well as the
corresponding rectification factors which clearly illustrate the asymmetry
between left and right propagation in such systems. We examine
not only the existence but also the dynamical stability of the plane wave 
states. Lastly, we generalize our numerical considerations to the more physically relevant
case of Gaussian initial wavepackets and confirm that the asymmetry in
the transmission properties also persists in the case of such wavepackets.
\end{abstract}

\maketitle

\section{Introduction} 

In the last two decades, the subject of nonlinear dynamical lattices
has gained considerable attraction and interest due to its emergence and relevance to a wide range
of diverse applications. These include, among
others, arrays of nonlinear-optical waveguides \cite{moti}, 
Bose-Einstein condensates (BECs) in periodic potentials \cite{ober},
micromechanical cantilever arrays~\cite{sievers}, 
Josephson-junction ladders \cite{alex},
granular crystals of beads interacting through Hertzian contacts~\cite{theo10}, layered antiferromagnetic crystals~\cite{lars3},
halide-bridged transition metal complexes~\cite{swanson},
 and dynamical models of the DNA double strand \cite{Peybi}. 

On the other hand, a specific phenomenon that has been 
intensely explored in a wide variety of recent
studies is that of 
potentially asymmetric (\textit{i.e}., non-reciprocal) wave propagation.
This has been examined e.g., in asymmetric phonon transmission through a 
nonlinear interface 
layer between two very dissimilar crystals \cite{kosevich}.
Another example is the proposed thermal diode~\cite{l1} which, in turn, led 
to its experimental realization ~\cite{l3}.  The optical diode was  
theoretically suggested in \cite{l5} (see also \cite{l7} for a setup of unidirectional transmission 
in photonics crystals) and experimentally achieved in \cite{l9}. Such rectification
effects have also been proposed in left-handed metamaterials~\cite{l10},
granular crystals~\cite{chiar} and systems with gain-loss bearing
so-called $\mathcal{PT}$-symmetry~\cite{kot4,D'A2012}, among others.

In the present work, we focus on an, arguably simpler, implementation
of the diode effect, which is fundamentally due to nonlinearity,
as has been presented recently in the work of~\cite{casati}.
There, a linear chain was considered with a pair (or more) of nonlinear
sites between the two linear ends of the chain. The nonlinear nature
of the dynamics, coupled to a potential asymmetry between the characteristics
of the nonlinear sites, was at the heart of the asymmetric propagation
observed. The nonlinear sites were modeled as
a prototypical system that has arisen in numerous 
applications, either as a direct model of relevance or as an 
envelope approximation in the form of the so-called
discrete nonlinear Schr{\"o}dinger (DNLS) equation~\cite{book}. 
While structurally simple, this model incorporates the
fundamental characteristics of such lattices, namely diffraction
(\textit{i.e}., a discrete analogue of dispersion) and nonlinearity.
The scenario of~\cite{casati} explores the standard
cubic nonlinearity associated with the Kerr effect~\cite{moti}.
However, often in applications, other types of nonlinearities are important
as well. For instance,  defocusing lithium
niobate waveguide arrays exhibit a different type of nonlinearity,
namely a saturable, defocusing one due to the photovoltaic effect \cite{kip}. 
In the latter context, dark solitons have been identified not only
in regular homogeneous lattices, but also in higher gaps~\cite{rongdong1},
as parts of multi-component soliton complexes (such as dark-bright
solitary waves)~\cite{rongdong2}, and even in heterogeneous
chains with alternating couplings~\cite{rongdong3}.

In the present work, we combine the experimentally accessible
form of the saturable nonlinearity with the asymmetric propagation
nonlinear phenomenology of~\cite{casati}. 
The model setup is presented in Section II.
We consider in Section III exact plane
wave solutions by solving the linear parts of the chain and gluing
them through the nonlinear saturable ``defect'' sites. In this
way, we identify the asymmetry between left and right transmittivities,
due to the nonlinear propagation through the asymmetric dimer, trimer
or more generally oligomer (\textit{i.e}., few site) configuration. We explore
the stability of these configurations and generically identify them
as unstable. The dynamical evolution of the corresponding instabilities
is explored through direct numerical simulations. Finally, more realistic
(for experimental purposes) wavepackets of a Gaussian form are also
considered in Section IV 
and the manifestation of the asymmetry in propagation
under such initial data is systematically quantified.
Section V summarizes our findings and presents our conclusions
including some possible directions of future work.

\section{The Model}

Motivated by the above application of lithium niobate waveguide
arrays, 
we consider a nonlinear Schr{\"o}dinger type chain with governing equation
\begin{equation} \label{tdnls}
i\dot\phi_n(t)+\phi_{n+1}(t)+\phi_{n-1}(t)=\frac{\gamma_n \phi_n(t)} {1+|\phi_n(t)|^2} 
\end{equation}
for $\gamma_n\in\mathds{R}$ and $\phi_n\in\mathds{C}$.   Here $t$ plays the role of the (spatial) evolution variable. The saturable nonlinearity term is present in a finite region in the middle of the chain.   That is, $\gamma_n\neq 0$ only for $1\leq n\leq N$.   The wave propagates freely 
\textit{i.e}., linearly outside of the finite region containing the nonlinearity.  The system is Hamiltonian \cite{Sam2013} with 
\begin{equation}\label{H}
\mathcal{H} = \displaystyle\sum_n \left[ \left( \phi_n^* \phi_{n+1} + \phi_n \phi_{n+1}^*\right) - \gamma_n \ln(1+|\phi_n|^2) \right].
\end{equation}
We will examine the transmission properties of stationary solutions that take the form of plane waves on the linear portions of the lattice.  We also explore the linear stability analysis 
for these solutions and for a number of dynamically unstable scenarios,
we evolve the corresponding solutions through direct numerical simulations
over the propagation parameter $t$.  Finally, we examine the propagation of more physically 
applicable localized Gaussian wavepackets and summarize the corresponding 
transmission properties in connection to the corresponding (potentially
observable experimentally) asymmetries.

\section{Stationary Solutions}
\label{secstat}

\subsection{Plane Waves}

\begin{figure}[tbp]
\begin{center}
\includegraphics[width=8cm,angle=0,clip]{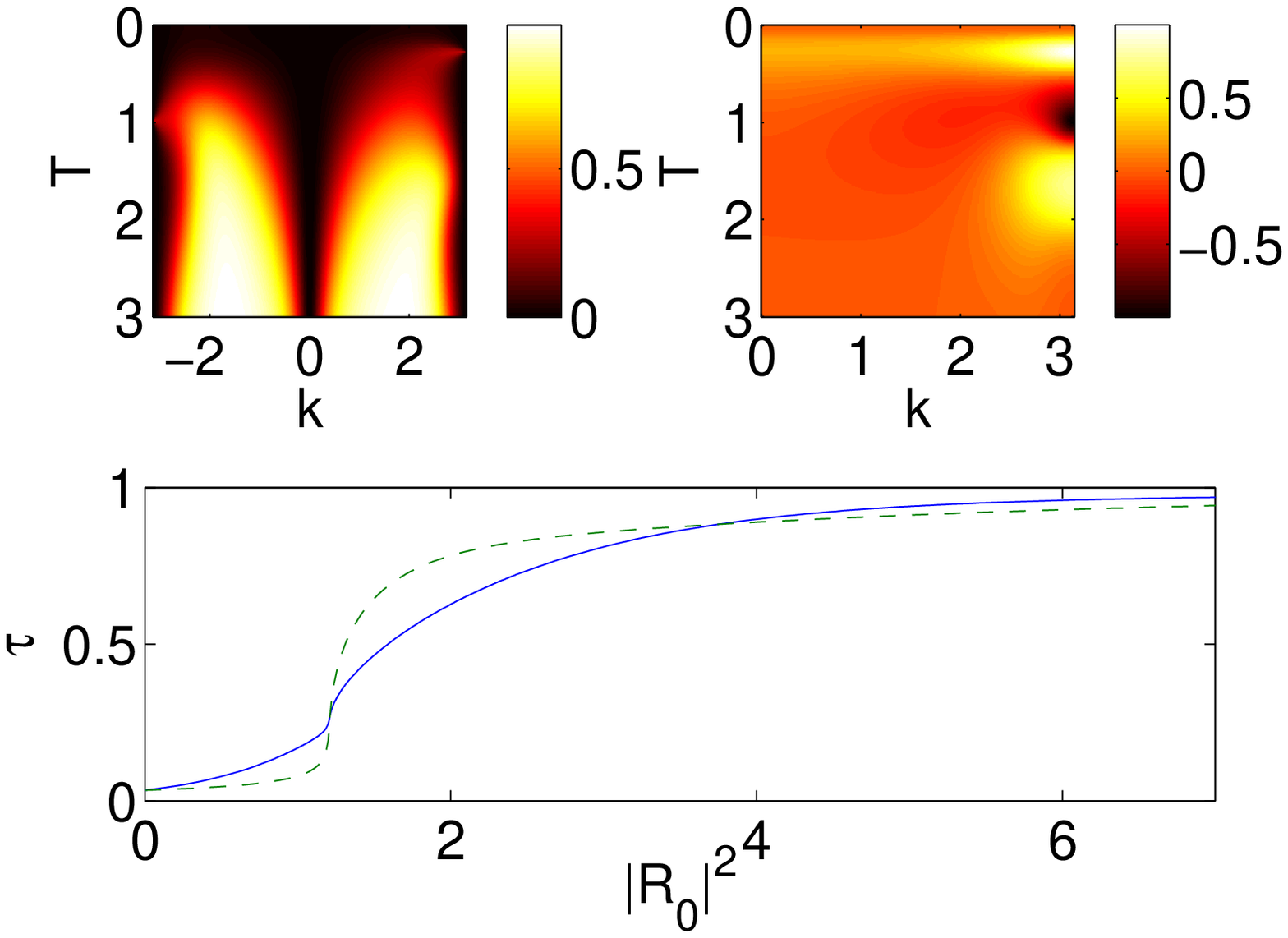}
\includegraphics[width=8cm,angle=0,clip]{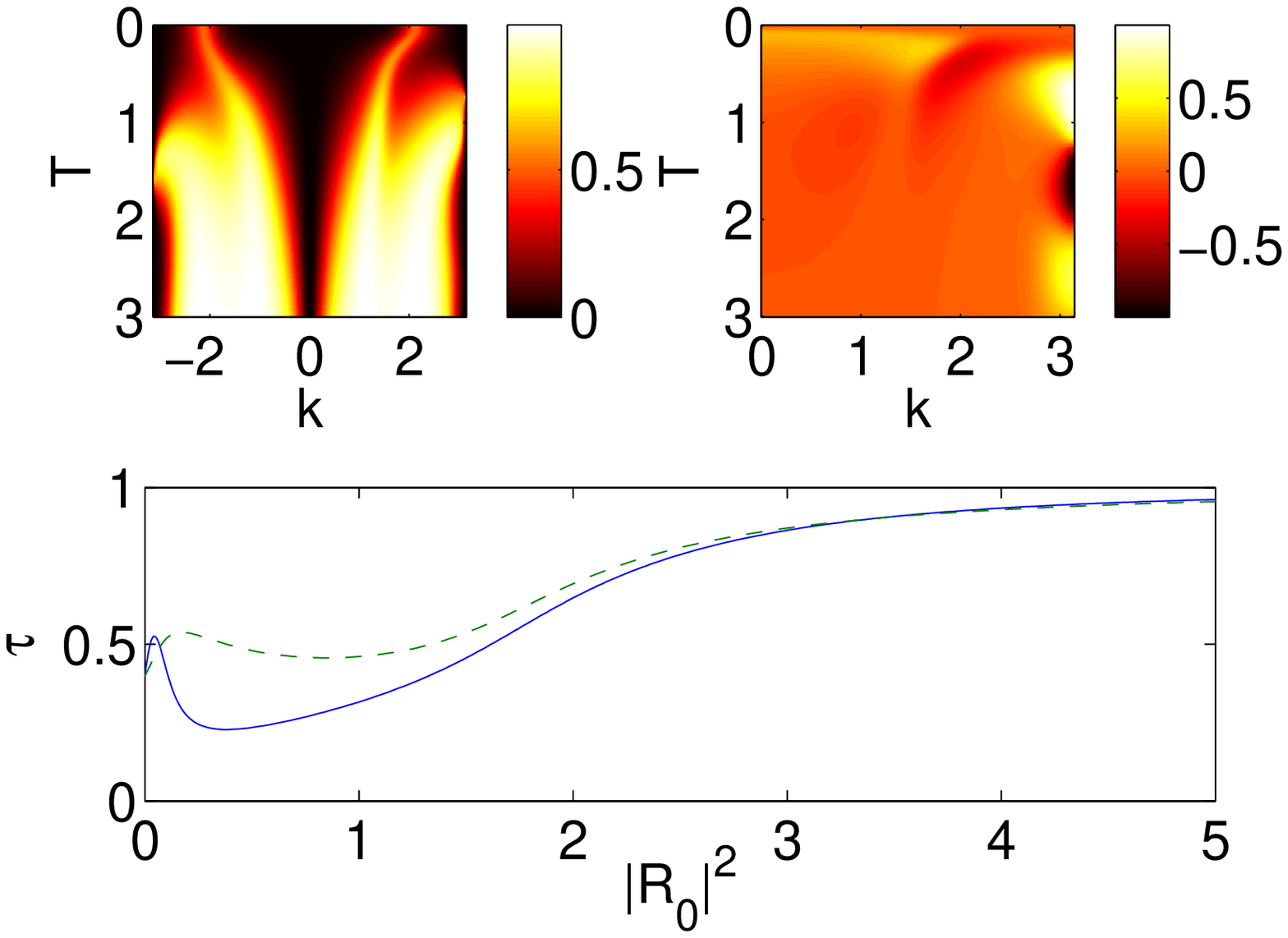}
\end{center}
\caption{
Each panel contains a contour plot of the transmission coefficient
$\tau(k,T)$ \mbox{({top left})} and a contour plot of the rectification factor 
$f(k,T)$ of Equation~(\ref{rectif}) ({top right}), plotted as a function of $k$ and $T$.  
The panels also show a typical example of the dependence of $\tau$ for $k=2$ (solid lines) and $k=-2$ (dashed lines), so as to illustrate the asymmetry between the propagation for left and right incident waves ({bottom} panel).  In the latter the dependence of $\tau$ is given as a function of $|R_0|^2$.  The left panel corresponds to $N=2$ and $\gamma_1=2.5$, $\gamma_2=5$ and the lattice size is $L = 200$ in this case.  The right panel is the $N=3$ case with $\gamma_1=1.5$, $\gamma_2=2$ and $\gamma_3=3.5$; here the lattice size is $L = 201$.   
}
\label{ftN=23}
\end{figure}

We seek standing wave solutions by setting $\phi_n = \psi_n e^{-i\omega t}$ for $\omega\in\mathds{R}$.  This gives a set of algebraic equations which can be written in the form of a backwards transfer map
\begin{equation}
\psi_{n-1} = -\psi_{n+1} + \left(-\omega + \frac{\gamma_n \psi_n} {1+|\psi_n|^2}\right)\psi_n 
\label{stat}
\end{equation}
for $\psi_n\in\mathds{C}$ independent of $t$.   Following the procedure in \cite{D'A2012} and for now assuming $k\geq 0$, we begin by assuming solutions in the form of plane waves on the linear portions of the chain.  That is,
\begin{equation}  \psi_n=\left\{ 
\begin{array}{ll}
R_0e^{ikn}+Re^{-ikn} & n\leq 1\\
Te^{ikn} & n\geq N
\end{array}
\right.
\label{plwv}  \end{equation}
with $R_0, R, T\in\mathds{C}$ representing the incident, reflected and transmitted amplitudes, respectively.  The solution (\ref{plwv}) solves Eq. (\ref{stat}) for $n\not\in\{1, \dots, N\}$ only if the wavenumber $k$ satisfies $\omega=-2 \cos(k)$.   Also directly from (\ref{plwv}) with $n=0, 1$ we have 
\begin{equation}R_0=\frac{e^{-ik}\psi_0-\psi_1}{e^{-ik}-e^{ik}} \qquad \mbox{ and } \qquad  R=\frac{e^{ik}\psi_0-\psi_1}{e^{ik}-e^{-ik}}. \label{R0R}\end{equation}
The stationary solution across the whole lattice is then known by the following procedure:  given values for $\gamma_n$  for $n\in \{1, \dots, N\}$ we start by specifying values for $k$ and $T$.  Then we compute $\psi_0, \psi_1, \dots \psi_{N-1}$ via Eq. (\ref{stat}) and then $R, R_0$ by (\ref{R0R}).  Such a procedure of finding the input as a function of the output is referred to as a ``fixed output problem'' \cite{knapp}.

Stationary solutions where the amplitude $R_0$ is incident from the right-hand-side and the wavenumber is taken as $-k \leq 0$ can also be formulated in a similar way.  In order to avoid swapping the format of \eqref{plwv} (so that $R_0, R$ would apply on the right and $T$ on left), it is more convenient to leave \eqref{plwv} as-is with {\it positive} wave number $k$ and instead flip left-to-right the configuration of the nonlinearities, \textit{i.e}. $\{\gamma_1, \gamma_2, \dots, \gamma_N\} \rightarrow \{ \gamma_N, \gamma_{N-1}, \dots, \gamma_1\}$.  In this way the computation of the solution for negative wavenumber is unchanged from the above outline aside from the swap of the order of the $\gamma$'s.  Plots of these plane wave stationary solutions are shown in the next section, where we also address the stability of their $t$-propagation.  In practice we truncate the lattice and refer to its finite length as $L$.

For a solution that has been determined by the processes described above, we next compute the transmission coefficient $\tau\stackrel{def.}{=} |T|^2/|R_0|^2$ explicitly assuming that $T$ is given.  For this purpose it is convenient to write $\psi_n\stackrel{def.}{=}Te^{ikN}\Psi_n$ with $n=N-l$ so that $l=0$ corresponds to $n=N$, and incrementing $l$ corresponds to decreasing $n$.  In this notation we have $\Psi_N = 1$ for $l=0$ and the value of $\Psi_n$ for each subsequent node towards the left is given by rewriting Eq. (\ref{stat}) as
\begin{equation}
\Psi_{N-l}=-\Psi_{N-l+2}+\delta_{N-l+1}\Psi_{N-l+1}\label{eq: alg}
\end{equation}
for $\delta_j\stackrel{def.}{=}-\omega+\frac{\gamma_j}{1+|T|^2|\Psi_j|^2}$.  See the Appendix where we record a few iterations of Eq. (\ref{eq: alg}).  Then by (\ref{R0R}) with $\psi_0, \psi_1$ computed according to Eq. (\ref{eq: alg}) we have 
\begin{equation} 
\begin{array}{lllllll}
R_0 = & { \frac{T}{ e^{-ik}-e^{ik}} }\left(   \delta_1-2e^{ik} \right) & \quad &
\tau= & \left| \frac{e^{ik}-e^{-ik}}{  2e^{ik}-\delta_1  }\right|^2 & \quad &
\mbox{ \ for \ } N=1\\
R_0 = &\frac{Te^{ik}}{ e^{-ik}-e^{ik}} \left(   -1+(\delta_1-e^{ik})(\delta_2-e^{ik}) \right) & \quad &
\tau=&\left| \frac{e^{ik}-e^{-ik}}{  1+(\delta_1-e^{ik})(e^{ik}-\delta_2)  }\right|^2 & \quad &
\mbox{ \ for \ } N=2
\end{array}
\end{equation}
and
\begin{eqnarray}
R_0 &=& \frac{Te^{2ik}}{ e^{-ik}-e^{ik}} \left(   \delta_1-e^{ik} + (\delta_3-e^{ik})(1-\delta_2(\delta_1-e^{ik}) )   \right)\nonumber\\
\tau &=& \left|  \frac{e^{ik}-e^{-ik}}{      e^{ik}-\delta_1 + (e^{ik}-\delta_3)(1-\delta_2(\delta_1-e^{ik}) )}\right|^2 \quad 
\mbox{ \ for \ } N=3.
\end{eqnarray}
In the linear case ($\gamma_1=\gamma_2=0$) and in the symmetric case ($\{\gamma_1, \gamma_2, \dots, \gamma_N\}  = \{ \gamma_N, \gamma_{N-1}, \dots, \gamma_1\}$ as an ordered set) it is immediately seen that $\tau$ is the same for waves incoming from the left or right side.  For $N=1$ the transmission is always symmetric.

We also define here a quantity to measure the asymmetric propagation.  We will use the definition of a rectification factor $f$ in the form of
\begin{eqnarray}
f = \frac{\tau(k,T) - \tau_{flip}(k, T)}{\tau(k,T) + \tau_{flip}(k, T)}
\label{rectif}
\end{eqnarray}
where the quantity $\tau(k,T)$ corresponds to transmission of a left-incoming wave with positive wavenumber $k\geq 0$ and $\tau_{flip}(k,T)$ with $k\geq 0$ is equivalent to the transmission of a right-incoming wave with negative wavenumber (recall the process described above of keeping $k$ positive while flipping the order of the $\gamma$'s).  This way nonzero values of $f$ in the range $[-1,1]$ measure the
asymmetry of transmission in the system.  Symmetry in transmission corresponds to $f=0$ and $f>0$ corresponds to greater transmission of incident waves originating from the left (transmitted on the right) as compared with incident waves originating from the right (transmitted on the left).  Of course, $f<0$ corresponds to greater transmission of waves originating from the right.

Figure \ref{ftN=23} shows plots of the transmission coefficient $\tau$ and the rectification factor $f$ as a function of the amplitude $T$ and the wavenumber $k$ of the extended plane wave solutions.    We find that whether more is transmitted for waves incoming from the right or left is variable as a function of $k$
and $T$.  Notice that values for $\gamma$'s are chosen in Figure \ref{ftN=23} to be such that $\gamma_1 < \gamma_2$ in the $N=2$ case and $\gamma_1< \gamma_2< \gamma_3$ in the $N=3$ case.  In other words, with increasing $\gamma$'s from left to right we observe that transmission properties vary with the choice of the parameters $T,k$.  We find that in accordance with our above analysis the $N=1$ case is symmetric.  Although we do not show an $N=1$ analogue of Figure \ref{ftN=23}, such plots look similar to Fig. \ref{ftN=23} but there is exact symmetry and $f=0$ for all $T,k$. It is interesting to point out here that the
rectification factor appears to acquire its largest (absolute)
values for $k$ close to $\pi$ \textit{i.e}., at the edge of the Brillouin
zone. Furthermore, both in the $N=2$ and in the $N=3$ case,
the dependence of $f$ near this value appears to be a non-sign-definite
function of $T$ (\textit{i.e}., different ranges of $T$ values appear to
favor propagation in one or the other direction).

\subsection{Stability}
\label{secstab}

\begin{figure}[tbp]
\begin{center}
\includegraphics[width=8cm,angle=0,clip]{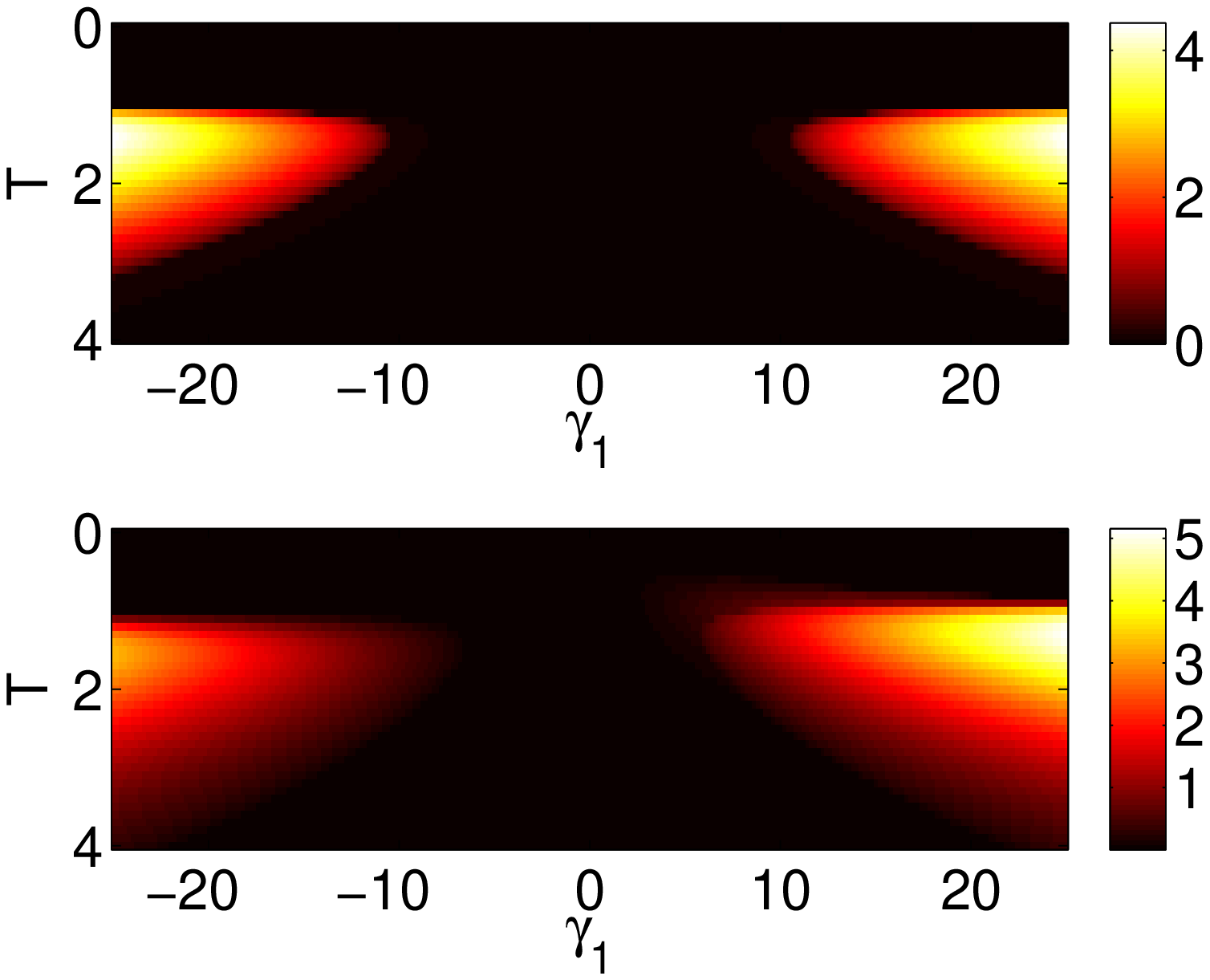}
\includegraphics[width=8cm,angle=0,clip]{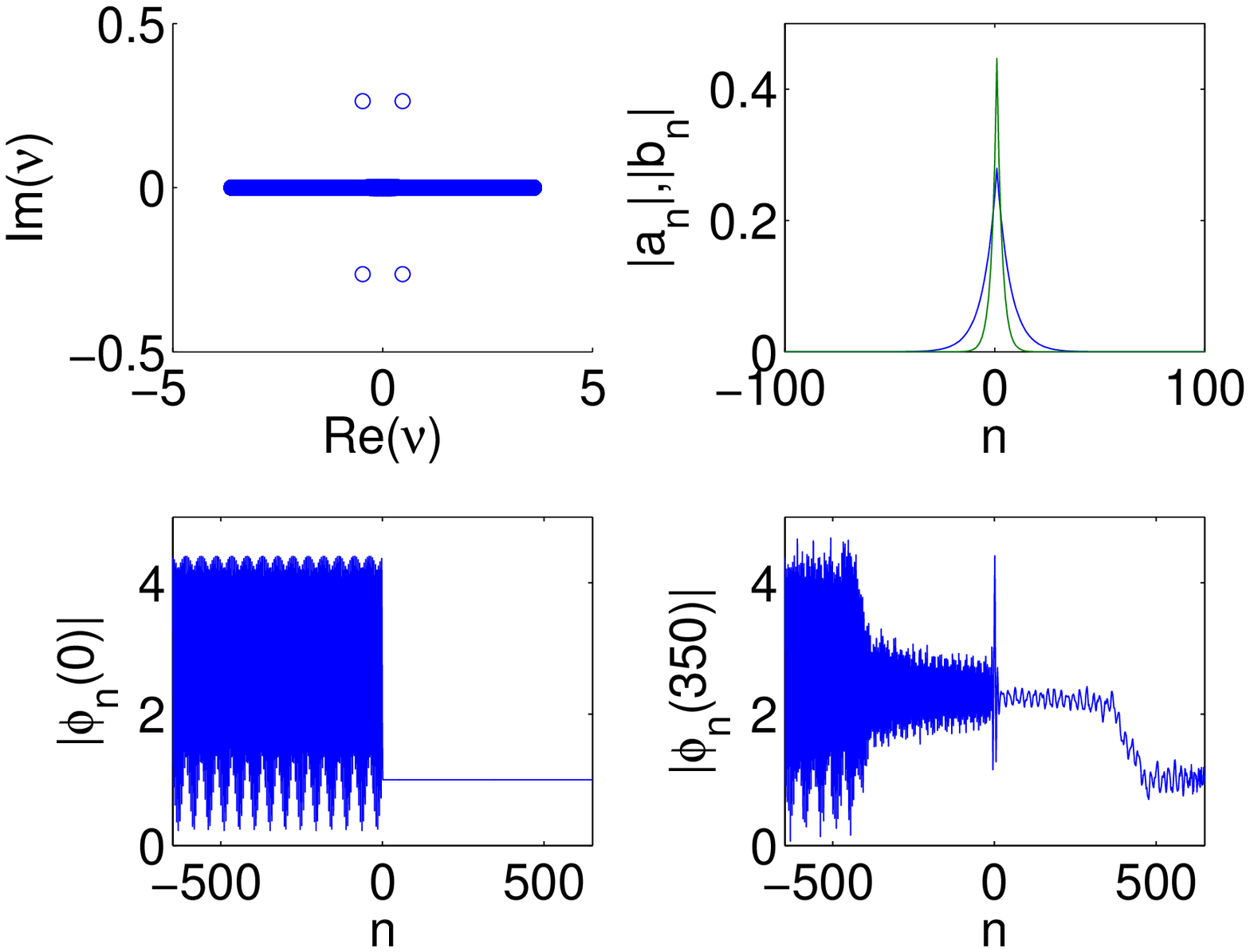}
\caption{
The left panel depicts the value of $|\min({\rm Im}(\nu)|$, for $N=1$, plotted as a function of $\gamma_1$ and $T$; the top graph shows $k = \pi/2$ and in the bottom graph $k = 2.5$.  The lattice length in the two left panel plots is $L=99$.  These two plots show that the magnitude of the minimum imaginary part of the calculated eigenvalue, \textit{i.e}. the strength of the instability, increases as the magnitude of $\gamma_{1}$ increases.  On the other hand for a fixed $\gamma_1$ value an extended solution of the form shown in Eq. \eqref{plwv} is tending toward stabilization for large $T$.  In the right panel we show four plots that correspond to a dim but nonzero region on the plot of $|\min({\rm Im}(\nu)|$ in the left panel.  That is, the right-hand four plots correspond to $k=2.5$, $\gamma_1=5$ and $T=1$. The four plots show the eigenvalues in the complex plane (top left), eigenvector magnitude (top right with $|a_n|$ blue and $|b_n|$ green), initial profile of the plane wave at $t=0$ (bottom left) and a later profile at $t=350$ of the plane wave (bottom right).    For this instability the eigenvalues are in the form of a quartet.
}
\label{timepropstabN=1}
\end{center}
\end{figure}

\begin{figure}[tbp]
\begin{center}
\includegraphics[width=8cm,angle=0,clip]{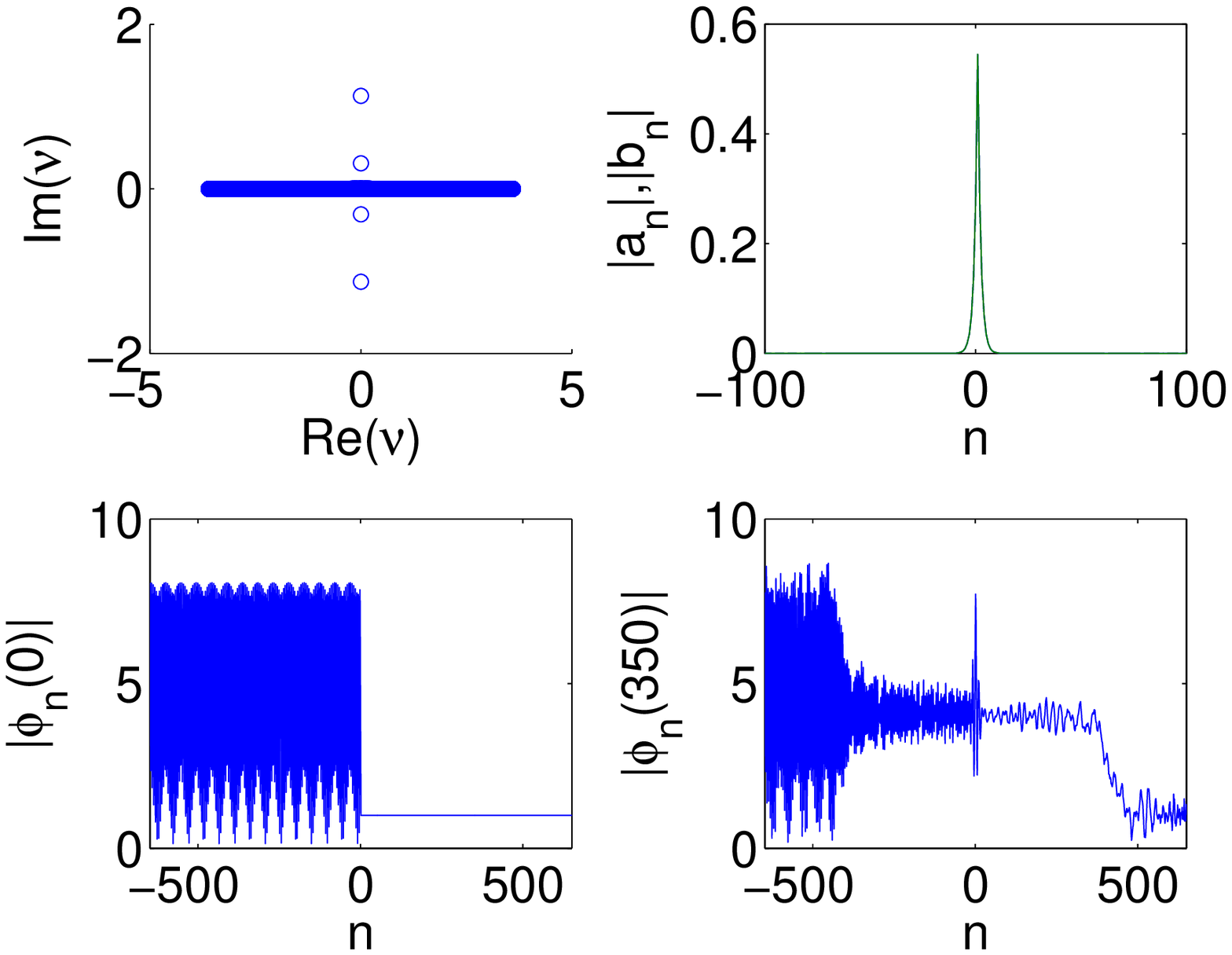}
\includegraphics[width=8cm,angle=0,clip]{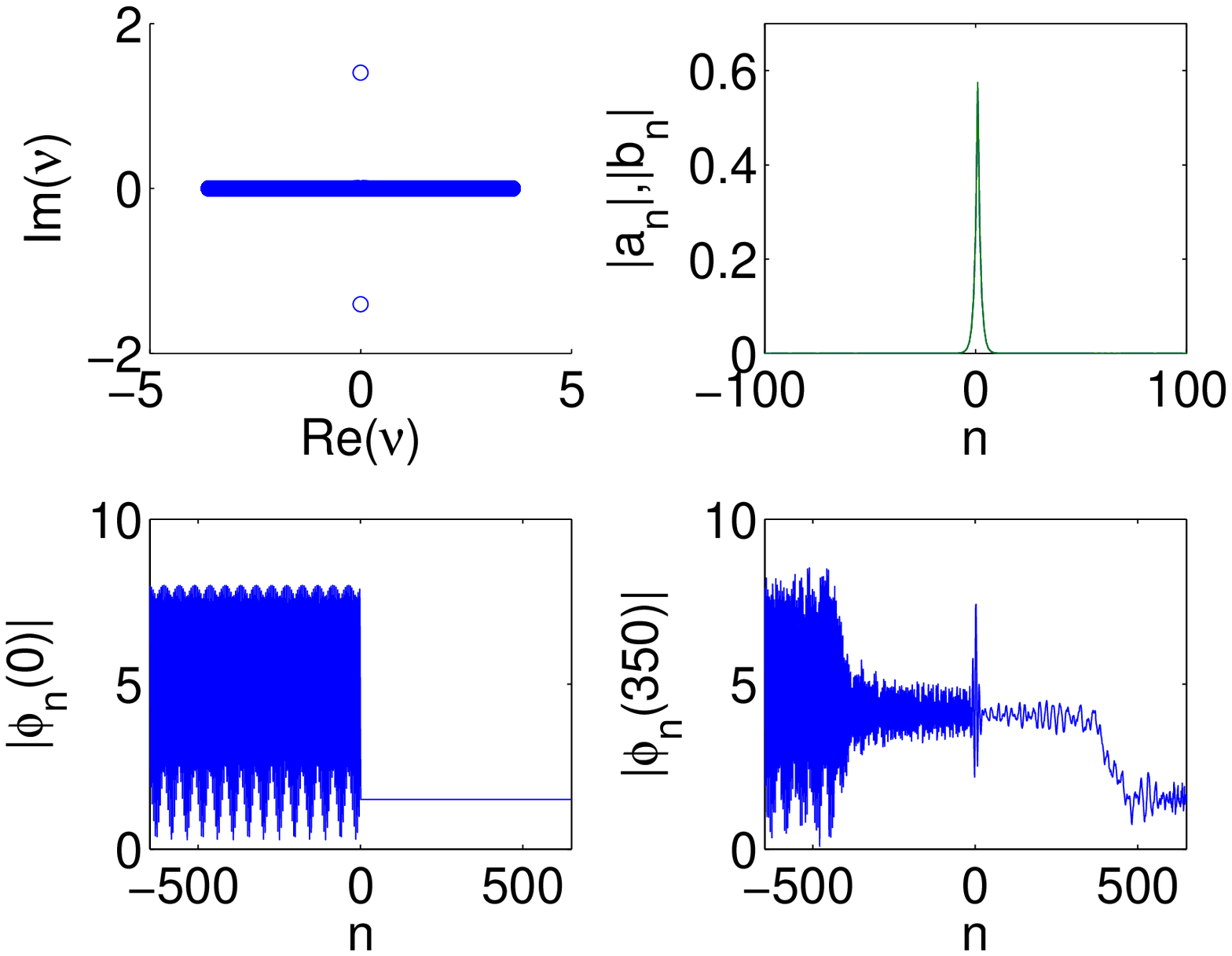}
\caption{
Here we focus on parameter values that correspond to a bright region in the left panel of Figure \ref{timepropstabN=1}.  We show four plots similar to the right panel of Figure \ref{timepropstabN=1}.  Here we have $N=1$, $L=1299$ and $k=2.5$.  The left panel of four plots corresponds to $\gamma_1 = 9.5$, $T=1$, and the right panel corresponds to $\gamma_1 = 10$, $T=1.5$.  Comparing the three sets of four plots in the present figure and in Figure \ref{timepropstabN=1} shows the transition in the eigenvalue plots as we move towards brighter regions of the $|\min({\rm Im}(\nu)|$ diagram.
}
\label{timepropN=1}
\end{center}
\end{figure}

\begin{figure}[tbp]
\begin{center}
\includegraphics[width=8cm,angle=0,clip]{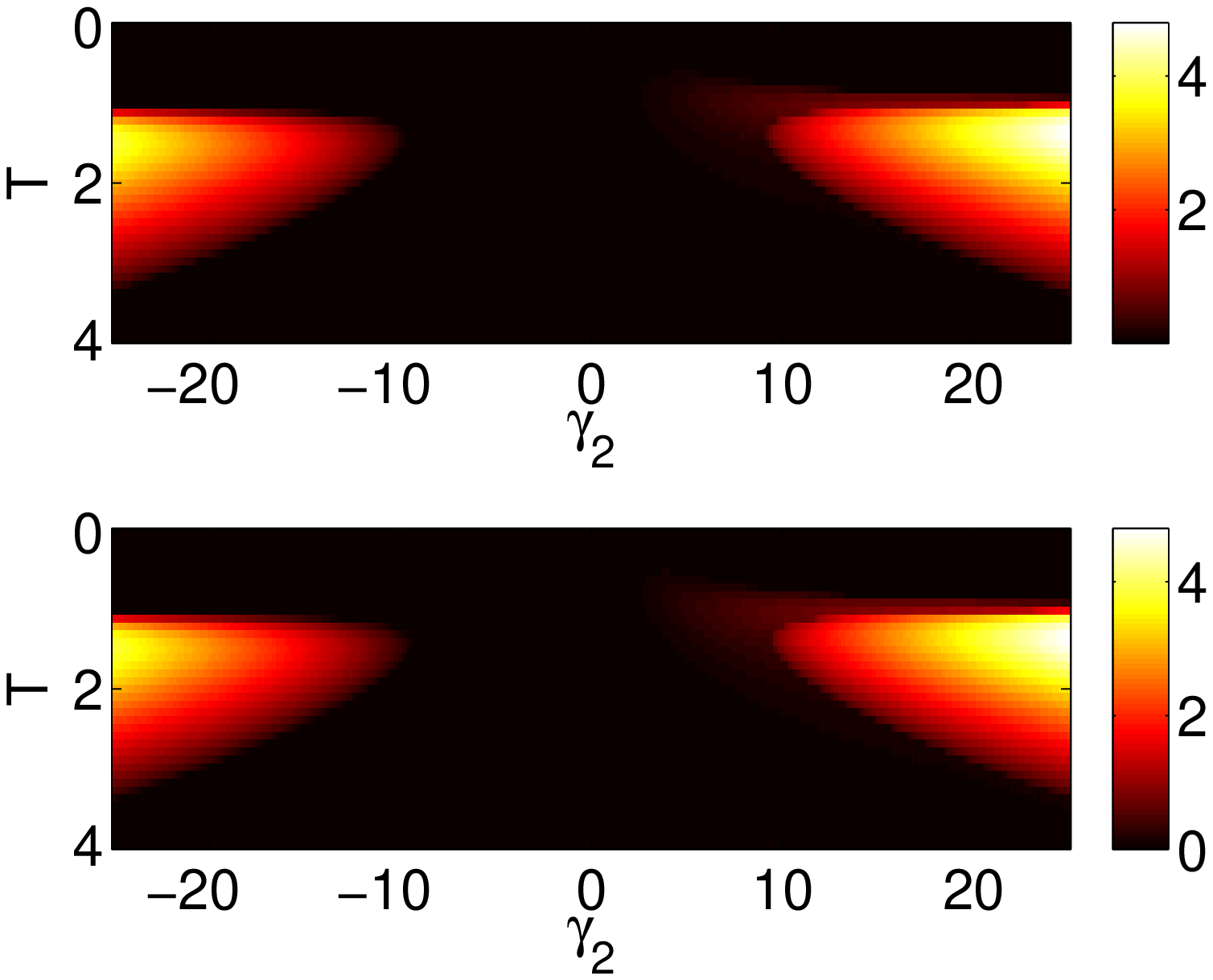}
\includegraphics[width=8cm,angle=0,clip]{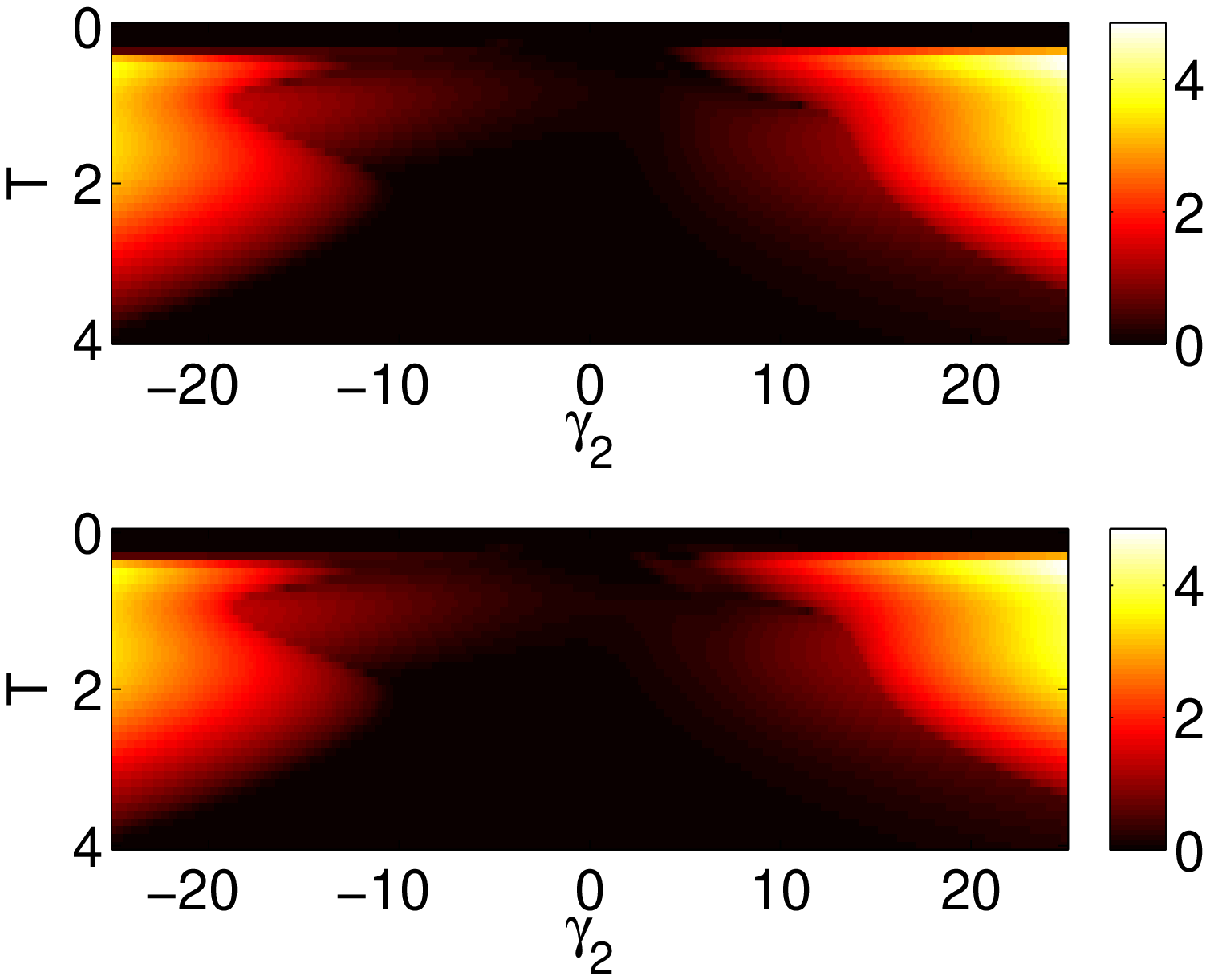}
\caption{
The plots are similar to the left panel in Fig. \ref{timepropstabN=1}.  Here the left panel corresponds to  $N=2$, $L=100$ and we plot $|min(Im(\nu)|$ as a function of $\gamma_2$ and $T$ while the value of $\gamma_1$ is fixed: $\gamma_1=1$ in the top graph and $\gamma_1=-1$ in the bottom graph.  Here the right panel corresponds to $N=3$, $L=101$ and we plot $|\min({\rm Im}(\nu)|$ as a function of $\gamma_2$ and $T$ while the values of $\gamma_1$ and $\gamma_3$ are fixed:  $\gamma_1=1$, $\gamma_3 = 5$ in the top graph and $\gamma_1=-1$,$\gamma_3 = 5$ in the bottom graph.
}
\label{stabN=23}
\end{center}
\end{figure}

\begin{figure}[tbp]
\begin{center}
\includegraphics[width=8cm,angle=0,clip]{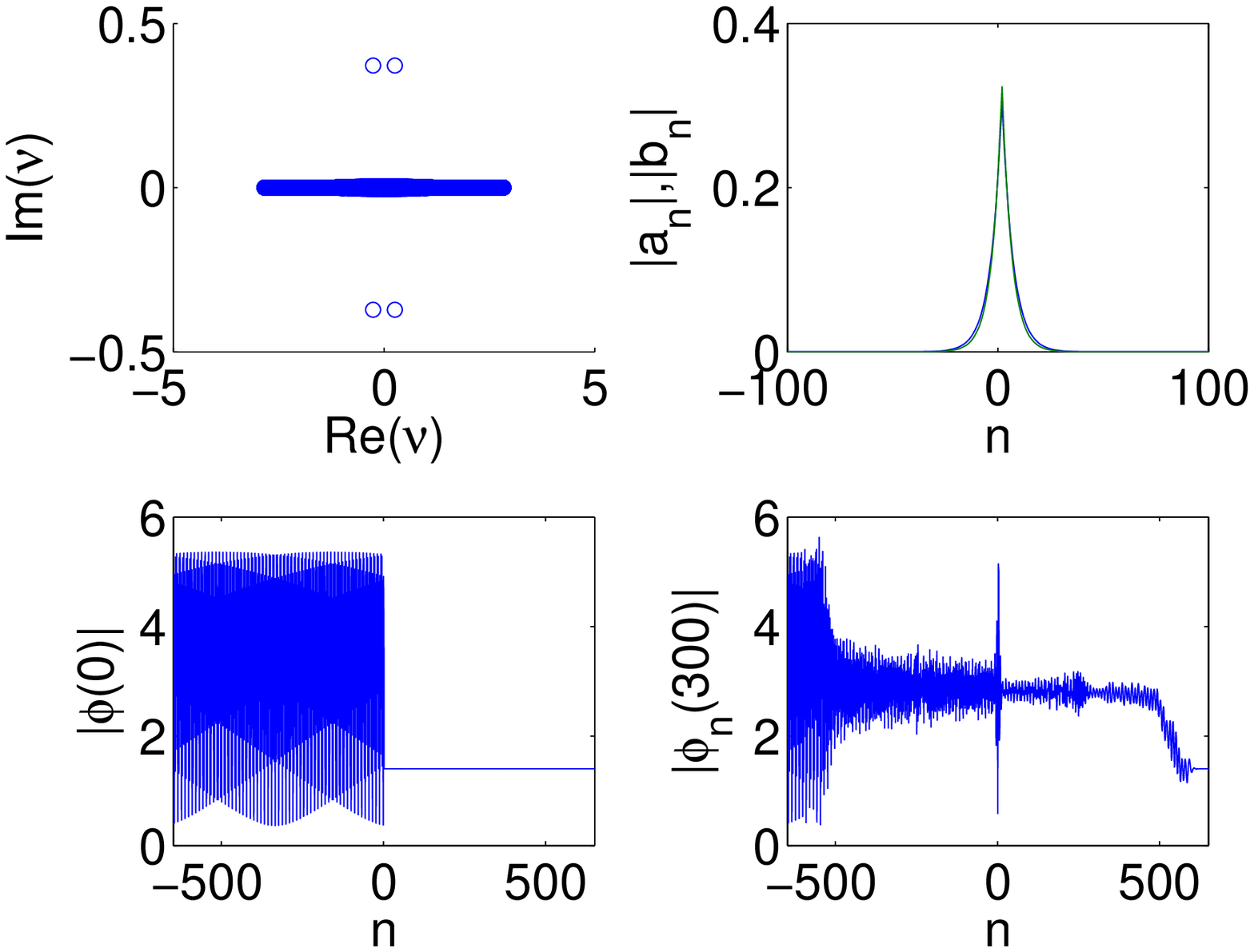}
\includegraphics[width=8cm,angle=0,clip]{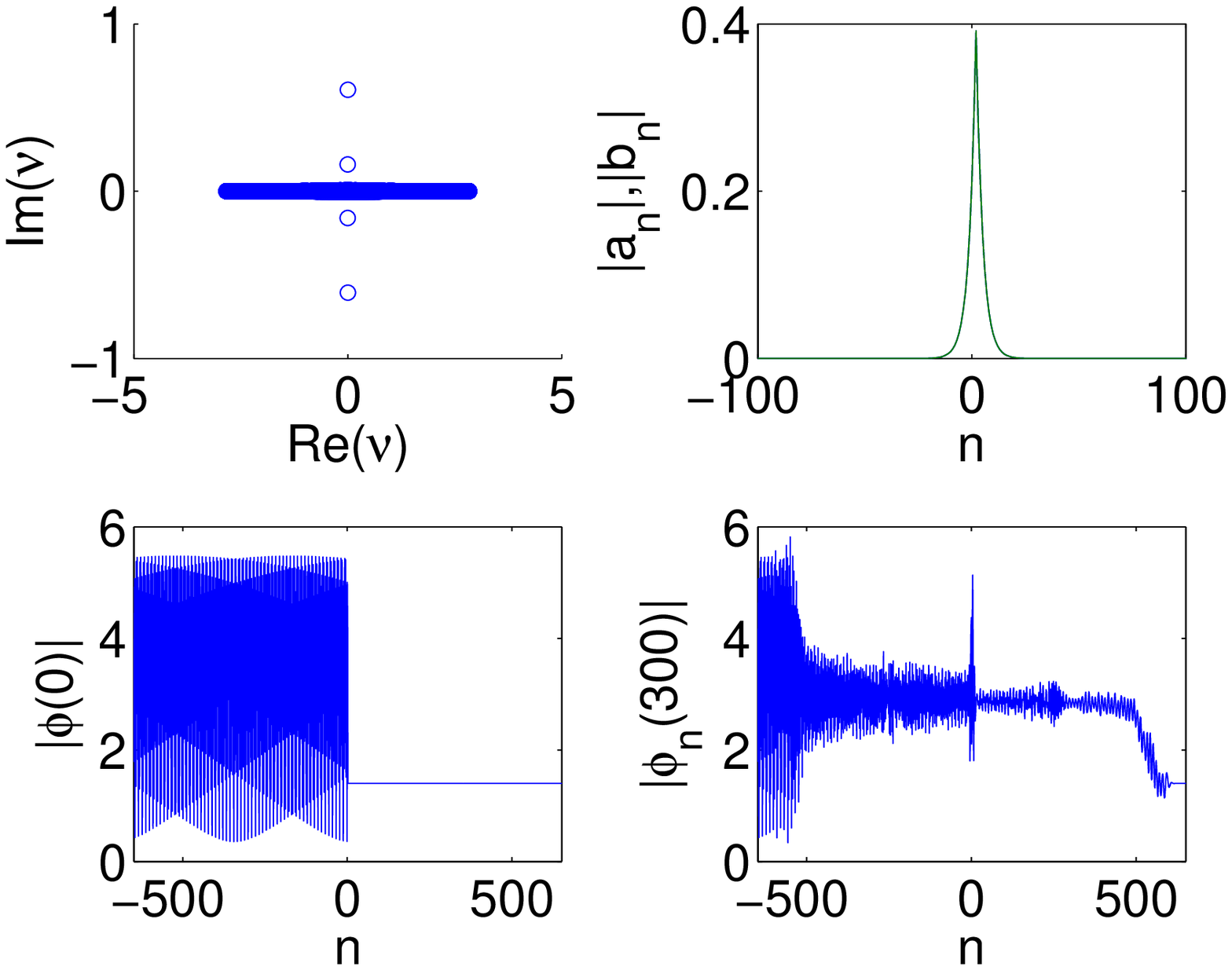}
\caption{
Here we focus on parameter values that correspond to the left panel of Figure \ref{stabN=23}.  Again we show four plots similar to the right panel of Figure \ref{timepropstabN=1}.  Here we have $N=2$, $L=1300$ and $k=2$.  The left panel of four plots corresponds to $\gamma_1 = -1, \gamma_2 = 9.25$, $T=1.4$, and the right panel corresponds to $\gamma_1 = -1, \gamma_2 = 9.5$, $T=1.4$.  Comparing these two sets of four plots shows the transition in the eigenvalue plots as we move towards brighter regions of the appropriate $|min(Im(\nu)|$ diagram in Figure \ref{stabN=23}.
}
\label{timepropN=2}
\end{center}
\end{figure}

\begin{figure}[tbp]
\begin{center}
\includegraphics[width=8cm,angle=0,clip]{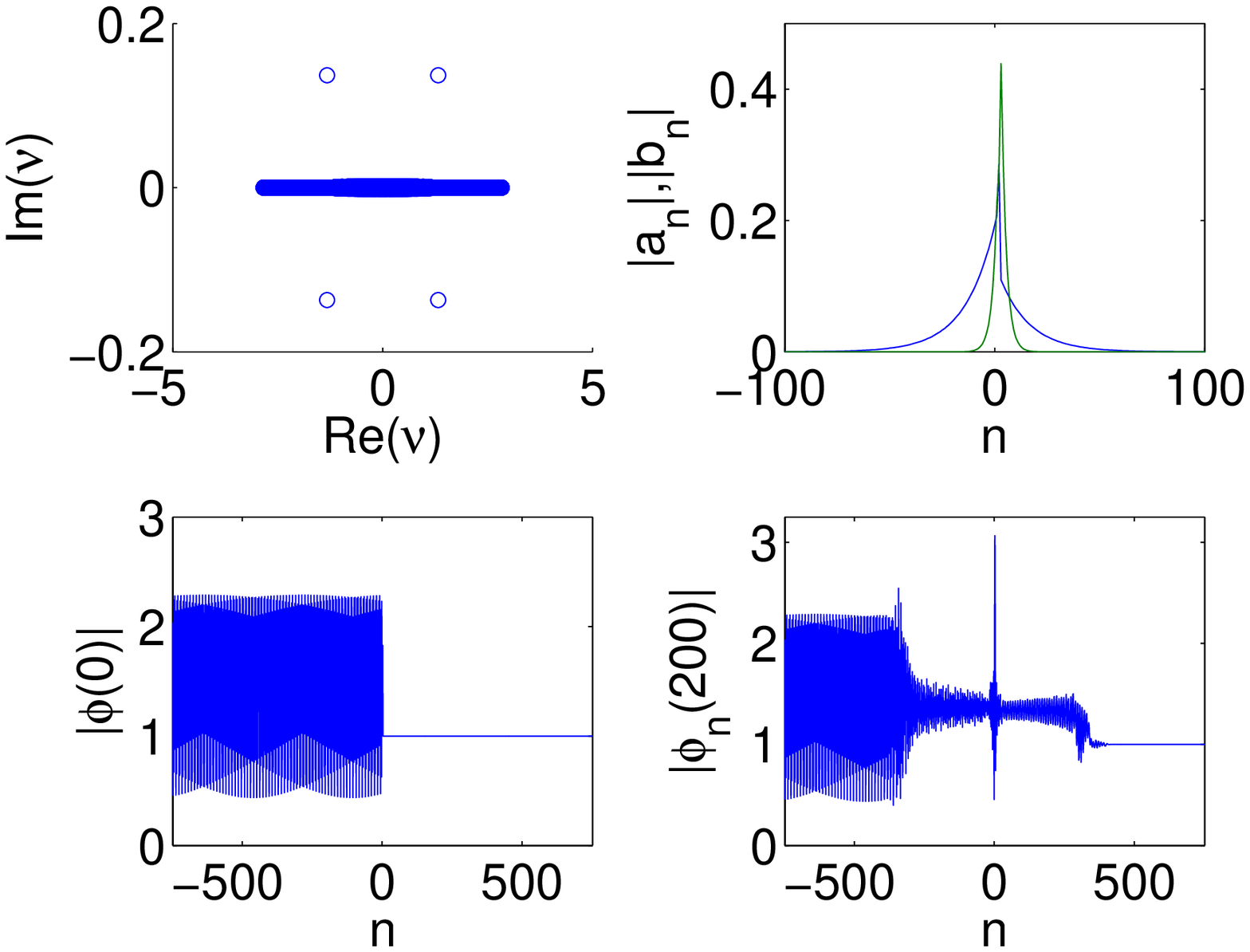}
\includegraphics[width=8cm,angle=0,clip]{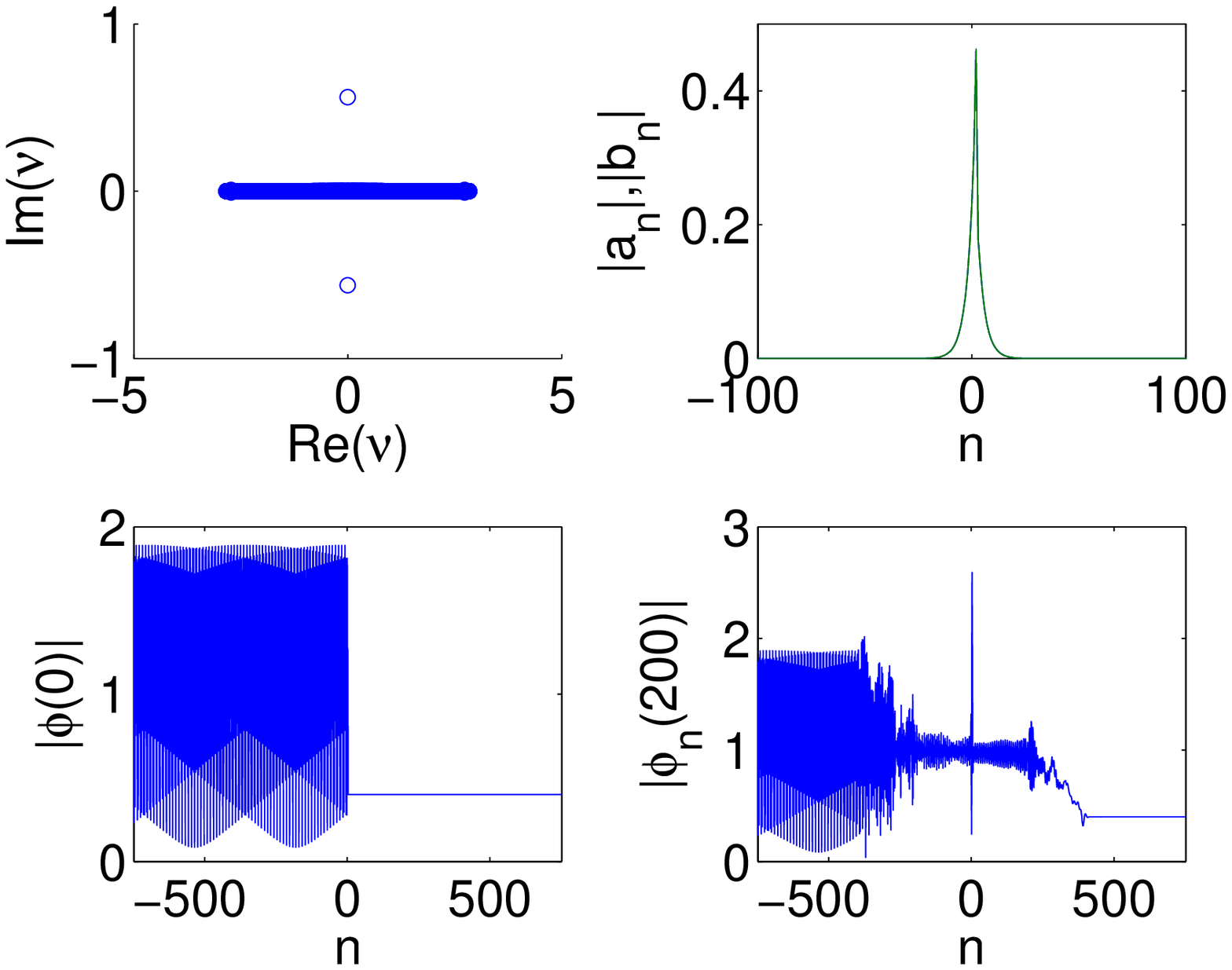}
\caption{
Here we focus on parameter values that correspond to the right panel of Figure \ref{stabN=23}.  Again we show four plots similar to the right panel of Figure \ref{timepropstabN=1}.  Here we have $N=3$, $L=1301$ and $k=2$.  The left panel of four plots corresponds to $\gamma_1 = 1, \gamma_2 = 3.25, \gamma_3 = 4$, $T=1$, and the right panel corresponds to $\gamma_1 = 1, \gamma_2 = 5.5, \gamma_3 = 4$, $T=0.4$.  Comparing these two sets of four plots shows the transition in the eigenvalue plots as we move towards brighter regions of the appropriate $|min(Im(\nu))|$ diagram in Figure \ref{stabN=23}.
}
\label{timepropN=3}
\end{center}
\end{figure}

\begin{figure}[tbp]
\begin{center}
\includegraphics[width=8cm,angle=0,clip]{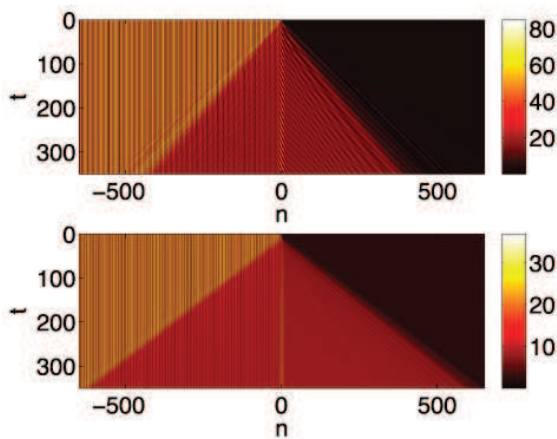}
\caption{
The plots show $|\phi_n(t)|^2$ as a function of $n$ and $t$.  The top plot corresponds to parameters the same as in the right four plots in Fig. \ref{timepropN=1}.  The bottom plot corresponds to parameters the same as in the left four plots in Fig. \ref{timepropN=2}.
}
\label{spctm}
\end{center}
\end{figure}

In order to analyze spectral stability of stationary states of the form discussed in the previous subsection we write
\begin{equation}
\phi_n(t) = e^{-i \omega t} \left( \psi_n + \varepsilon \left( a_ne^{i\nu t}+b_n e^{-i\nu^* t} \right)  \right)
\label{pert}
\end{equation}
for $a_n, b_n,\nu \in\mathds{C}$, $\varepsilon$ small, and with $\psi_n$ being a stationary solution from the previous section.  The resultng linear stability equations then read
 \begin{equation}\label{mat}
 \nu \left(  \begin{array}{c} a_n\\ b_n^* \end{array}  \right)
 =
 \left( \begin{array}{cc} F_1 & F_2\\ F_3 & F_4 \end{array} \right)
\left(  \begin{array}{c} a_n\\ b_n^* \end{array}  \right)
\end{equation}
for
\begin{eqnarray}
& F_1 = diag\displaystyle\left(\omega-\frac{\gamma_n}{(1+|\psi_n|^2)^2}\right)+G, \qquad 
& F_2 = diag\displaystyle\left( \frac{\gamma_n \psi_n^2}{(1+|\psi_n|^2)^2}\right)\nonumber\\
& F_3 = diag\displaystyle\left(\frac{-\gamma_n\psi_n^2}{(1+|\psi_n|^2)^2}\right), \qquad 
& F_4 = diag\displaystyle\left(-\omega+\frac{\gamma_n}{(1+|\psi_n|^2)^2}\right)-G \label{F}
\end{eqnarray}
where $G$ is a sparse matrix with ones on both the super- and sub-diagonals.  Given a stationary plane wave solution $\psi_n$ and values of $\gamma_n$ which encode the nonlinearity for $1\leq n\leq N$, one then calculates the eigenvalues $\nu$ in (\ref{mat}).  If $\nu$ has a negative imaginary part this indicates that the perturbed solution $\phi_n(t)$ is unstable, as is easily seen by Eq. (\ref{pert}).  In practice, one diagonalizes a finite truncation of the matrix in (\ref{mat}), ensuring that the relevant eigenvalues are not affected by the truncation error.   In other words, $F_1, F_2, F_3, F_4$ and $G$ are all $L\times L$ matrices and in the matrix Eq. (\ref{mat}) it is now convenient to think of $a_n$ and $b_n$ as length $L$ column vectors. Furthermore, the Hamiltonian symmetry of the
solution ensures that the relevant instability eigenvalues come
either in pairs (if $\nu$ is imaginary) or in quartets (if $\nu$
is genuinely complex).

In Figures \ref{timepropstabN=1}, \ref{stabN=23} we show a plot of 
$\min({\rm Im}(\nu))$ as a function of $T$ and $\gamma$.  We find that an increase in the magnitude of a $\gamma_i$ parameter (with other nearby $\gamma$'s held fixed) leads to $\min({\rm Im}(\nu))$ of larger magnitude indicating greater instability.  Figures \ref{timepropstabN=1}, \ref{timepropN=1}, \ref{timepropN=2}, \ref{timepropN=3} show eigenvector and eigenvalue plots alongside snapshots of $\phi_n(t)$ to show the behaviour of typical propagation in the $t$ variable of the unstable plane waves.  The boundary conditions are calculated according to (\ref{plwv}) at $t=0$ and evolved by multiplying by $e^{-i\omega t}$ for $t>0$ so as to conform with (\ref{pert}).   The unstable plane wave solution, when propagated in the evolution variable, exhibits a few effects:  if $k>0$ then amplitude leaks over to the right-hand side (to the left if $k<0$), and due to the localized instability a peak appears in the center of the lattice.  Of course, given the conservation laws
of the system, the power $\sum_n |\phi_n(t)|^2$ and the Hamiltonian $\mathcal{H}(t)$ are preserved over $t$.  The figures also show a transition in the eigenvalue plots for unstable solutions.  A 
weak instability (corresponding to dim but nonzero regions of the $min(Im(\nu))$ plots of Figures \ref{timepropstabN=1}, \ref{stabN=23}) results in eigenvalue plots in the complex plane where a quartet appears off the real axis; see Figures \ref{timepropstabN=1} and \ref{timepropN=2}.  As the instability is
enhanced for larger values of $\gamma$  (comparably brighter regions of the $min(Im(\nu))$ plots), the two pairs constituting the quartet merges on the imaginary axis and subsequently split with one pair headed towards zero; see Figures \ref{timepropN=1}, \ref{timepropN=2}.  For the highest magnitude of instability (brightest regions on the plots of $min(Im(\nu))$) the eigenvalues indicating the instability are in the form of a pair on the imaginary axis; see Figures \ref{timepropN=1}, \ref{timepropN=3}. In the examples shown, the instability generically appears to transport power to the right part of the lattice,
deforming (decreasing the power of) the corresponding $n<0$
portion of the plane wave. On the other hand, critically (per the 
localized eigenvector of the instability), a localized mode
appears to form at the central nonlinear nodes within the domain.

In comparing our results with that of \cite{casati}, we find that the asymmetry associated with the saturable nonlinearity (presented here) is less pronounced to that of a system with a cubic nonlinearity and a linear potential term (presented in \cite{casati}).   In the $t$ propagation of extended solutions we can also compare the top plot in Fig. 3 of \cite{casati} with our Fig. \ref{spctm} in which we show plots over space and the propagation parameter.  The two systems both experience a concentration of amplitude at the center of the lattice as $t$ moves forward.  In the case of the cubic nonlinearity in \cite{casati} there are three concentrations of amplitude (two of which are moving).  Here we see only the one central concentration of amplitude at the center while sites nearby the center drop amplitude in comparison to the highest points of the initialized state at $t=0$.  Also, in contrast to \cite{casati} where the amplitude concentrations more dramatically rise above the background, here the central concentration of amplitude is more similar to the maximal amplitude of the initialized state.
 
\section{Propagation of a Gaussian}
\label{sec: Gaus}

\begin{figure}[tbp]
\begin{center}
\includegraphics[width=8cm,angle=0,clip]{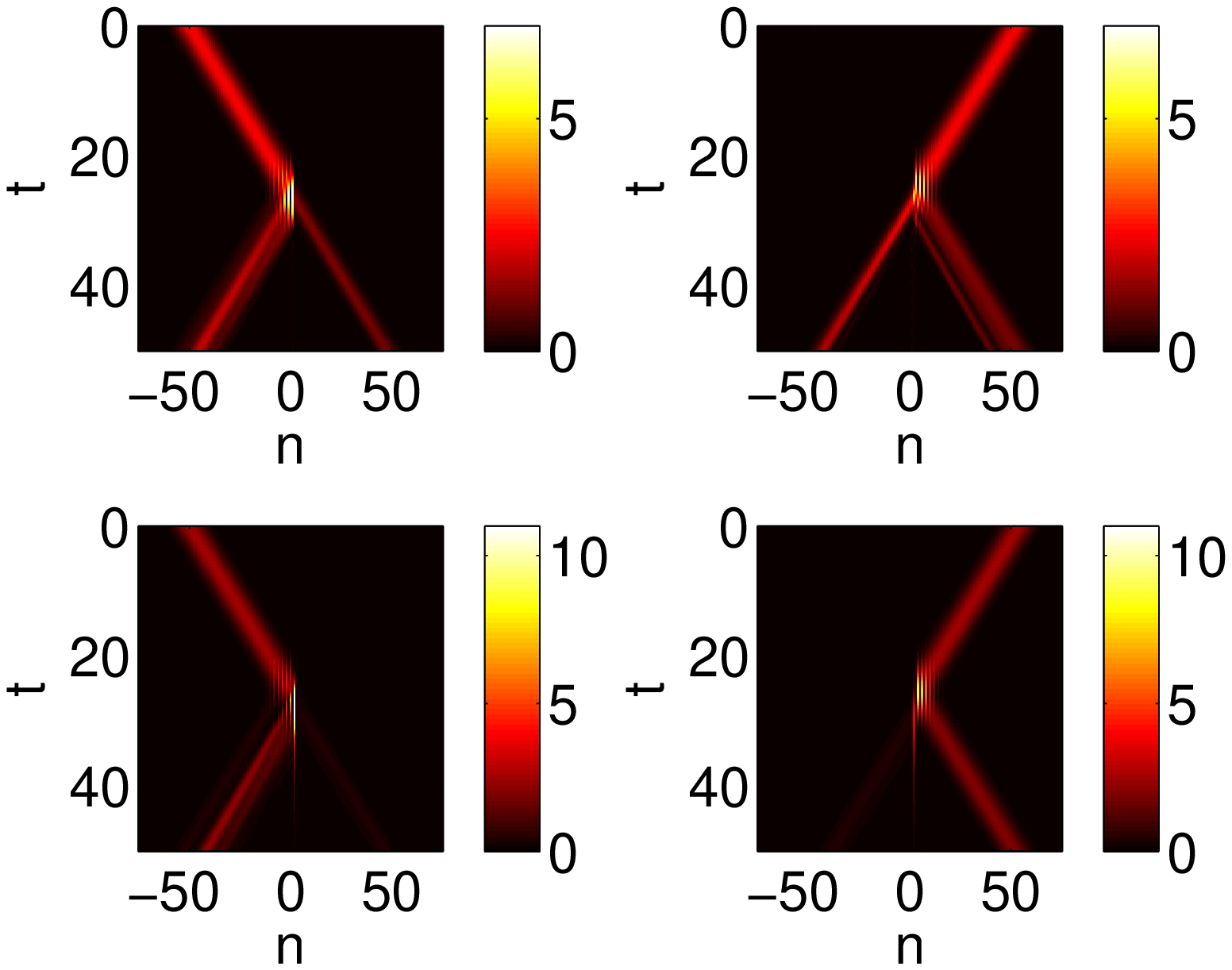}
\includegraphics[width=8cm,angle=0,clip]{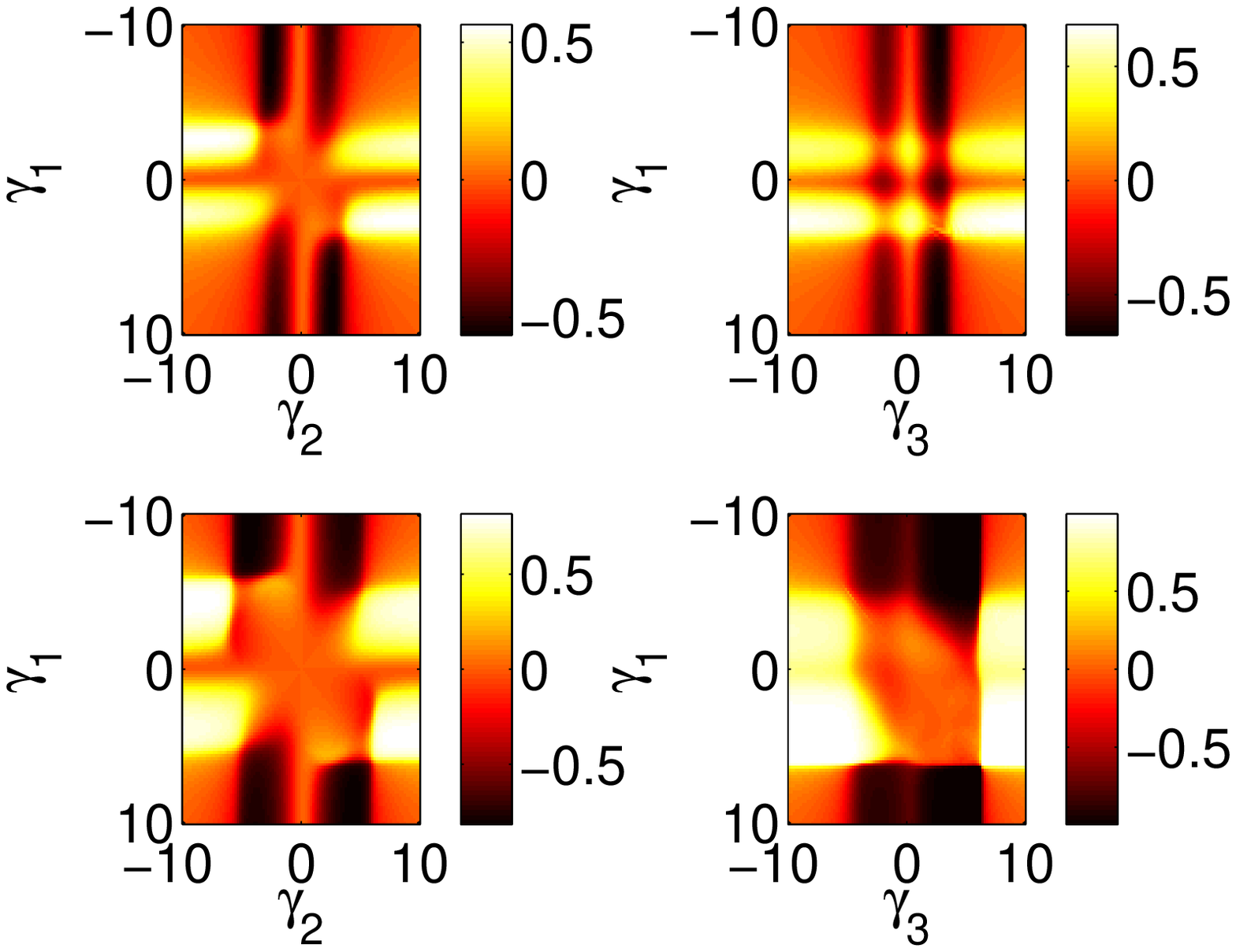}
\caption{
In the left four graphs, we plot $|\phi_n(t)|^2$ as a function of the lattice $n$ and the propagation variable $t$ for initial profile according to (\ref{gaus}) with $|I|^2=2.5$.  The left columns have wavenumber $k = \pi/2$ and starting position $n_0=-50$ and the right columns have wavenumber $k = -\pi/2$ and starting position $n_0 = 50$.  The top two plots correspond to $N=2$ with $\gamma_1=2.75$ and $\gamma_2=5$; we calculate that the rectification factor in this case is $f\approx -0.1767$. 
The bottom two plots correspond to $N=3$ with $\gamma_1=3$, $\gamma_2=3.25$, $\gamma_3=4$; we calculate that $f\approx -0.1525$. 
In the right four graphs, we plot the rectification factor of the Gaussian waves for $N=2$ (left column) and $N=3$ (right column) as a function of variable $\gamma$ parameters.  The top row corresponds to $|I|^2=1$ in (\ref{gaus}) and the bottom row to $|I|^2=2.5$.  For the $N=3$ plots, the value $\gamma_2=5$ (top) and $\gamma_2=3.25$ (bottom)  is fixed.
}
\label{gauss}
\end{center}
\end{figure}

Finally, in this section we look at the propagation of a Gaussian wavepacket through 
the lattice for each of the cases $N=1,2,3$. 
While it is less straightforward to prepare the delocalized 
initial conditions needed for the plane wave solutions of the previous
section (whose asymmetric propagation, however, can be analytically
quantified), preparing the Gaussian initial data of the present
section appears to be considerably more tractable in optical experiments
e.g., with lithium niobate waveguide arrays. On the other hand, in this
latter setting, we will have to rely on detailed numerical computations
of the rectification factor, as this set of initial conditions is less
amenable to detailed analytical considerations.
The wavepacket considered is given by the equation
\begin{equation}
\phi_n(0) = Ie^{-(n-n_0)^2/s^2}
\label{gaus}
\end{equation}
for starting position index $n_0$ and width parameter $s$. We measure the transmission at some value $t=t_0$ sufficiently large so that the wavepacket has interacted with the nonlinear region and, as a result, 
some portion of it has been accordingly transmitted through and reflected
from the relevant interval.  The transmission is then measured by
\begin{equation}
\tau_+ = \frac{ \sum_{n>N}|\phi_n(t_0)|^2 }{ \sum_n|\phi_n(t_0)|^2 } \qquad 
\tau_- = \frac{ \sum_{n<1}|\phi_n(t_0)|^2 }{ \sum_n|\phi_n(t_0)|^2 }
\end{equation}
for $k>0$ and $k<0$, respectively.  Then the rectification factor takes the form $f = (\tau_+ - \tau_-)/(\tau_+ + \tau_-)$.  The transmission is, of course, equal in both directions ($f=0$) in the $N=1$ case.  In Figure \ref{gauss} we plot $f$ as a function of $\gamma_1, \gamma_2$ in the case of $N=2$, and as a function of $\gamma_1, \gamma_3$ with $\gamma_2$ fixed in the case of $N=3$.  In Figure \ref{gauss} we also show some typical space-$t$ propagation plots for 
the Gaussian initial data case.  Similar to the plane wave solutions case, the rectification factor may be positive or negative as the parameters change.  Here the Gaussian in this particular case with $k=\pi/2$ has the following property.  For parameter values concentrated in the vertical and horizontal bands in the right-side four plots of Fig. \ref{gauss}  more is transmitted if the wave hits the {\it lower} $\gamma$-value first, in comparison to the wave hitting the larger $\gamma$-value first, \textit{i.e}., encountering the region which is closer to linear is more 
conducive towards transmission, while encountering the more nonlinear
sites at first is more prone to reflection, a feature that seems to
be intuitively justified.

\section{Conclusions}

In the present work, we considered a lattice setting where embedded
in a linear Schr{\"o}dinger chain was a nonlinear ``segment'' of the 
saturable type. Our analytical considerations
were focused around plane waves enabling us to analytically compute
both the transmittivity and the rectification factor between left-
and right-propagating such waves. These features evidence
the asymmetric nature of the propagation in a way that is
analytically tractable. This asymmetry can be explicitly
traced in the nonlinearity of the relevant setup. We also considered
the spectral stability of such states in which we observe some effects of 
low versus high rates of instability.  Finally, we considered the asymmetry 
of propagation of a Gaussian wavepacket. The latter was also clearly
evidenced both in the case of a dimer, as well as in that of a trimer,
paving the way for the experimental observation of relevant 
phenomena, such as the enhanced transmission of a wavepacket when
encountering a region of increasing, rather than that of a decreasing
nonlinear index profile.

There are numerous aspects that may be worthwhile to further explore.
In the context of lithium niobate waveguide arrays, it may be relevant
to examine settings that involve a genuinely nonlinear lattice but
with its central sites bearing a different nonlinearity than the
background. This type of ``spatial profile'' of the nonlinearity 
coefficient has attracted considerable interest in numerous
recent studies as evidenced by the review of~\cite{kartashov} and
may also be quite experimentally tractable. On the other hand,
the vast majority of the present studies on the nonlinearity-induced
asymmetry that we are aware of have focused chiefly on one-dimensional
configurations. However, it would be both more numerically challenging
and also theoretically intriguing to explore scattering of two-dimensional
wavepackets from a central (two-dimensional) segment of a lattice
which is genuinely nonlinear. Such aspects are currently under
investigation and will be reported in future publications.

Finally, we should note that after submission of this manuscript, we were notified of a related work in~\cite{assun}.  While the models analyzed in these works are fairly similar, distinctive features of the present work are (a) that we explored systematically the stability of the extended waves and we studied the outcomes of their dynamical instabilities; and (b) we also considered the effect of the incidence of a Gaussian wave packet.  Lastly, (c) although the latter work is restricted to dimers, our considerations here have been provided for general $N$ and examined for $N=1, 2, 3$.

\section{Acknowledgements}

 The authors acknowledge useful discussions with Stefano Lepri on the subject.  P.G.K acknowledges support from the National Science Foundation under grants CMMI-1000337, DMS-1312856, from ERC and FP7-People under the grant IRSES-606096 and from the US-AFOSR under grant FA9550-12-10332, as well as from the Binational Science Foundation under grant 2010239.

\appendix

\section{Appendix}
We show a few iterations of the backwards transfer map in (\ref{eq: alg}).  
\begin{eqnarray}
\label{eq: algN}
\delta_{N} &=& -\omega+\frac{\gamma_j}{1+|T|^2}  \nonumber\\
\Psi_{N-1} &=& - e^{ik}+\delta_N\nonumber\\
\delta_{N-1} &=& -\omega+\frac{\gamma_{N-1}}{1+|T|^2|\delta_N-e^{ik}|^2}\nonumber\\
\Psi_{N-2} &=&  - 1 + \delta_{N-1}(\delta_N-e^{ik})\nonumber\\
\delta_{N-2} &=& -\omega+\frac{\gamma_{N-2}}{1+|T|^2| 1 + \delta_{N-1}(e^{ik}-\delta_N)|^2}\nonumber\\
\Psi_{N-3} &=&  -\delta_{N-2}+(e^{ik}-\delta_N)(1-\delta_{N-2}\delta_{N-1})\nonumber\\
\delta_{N-3} &=& -\omega+\frac{\gamma_{N-3}}{1+|T|^2|\delta_{N-2}+(\delta_N-e^{ik})(1-\delta_{N-2}\delta_{N-1})|^2} \nonumber\\
\Psi_{N-4} &=& 1 -\delta_{N-3}\delta_{N-2}  + (e^{ik}-\delta_N)\left(   \delta_{N-1}+\delta_{N-3} (1-\delta_{N-2}\delta_{N-1})      \right) \label{dels}
\end{eqnarray}

\end{document}